\documentclass[preprint,showpacs,preprintnumbers,amsmath,amssymb]{revtex4}

\usepackage{graphicx}% Include figure files
\usepackage{dcolumn}% Align table columns on decimal point
\usepackage{bm}% bold math

\begin{document}

\title{Absence of a Diffusion Anomaly in Water Perpendicular to Hydrophobic Nanoconfining Walls}

\author{Sungho Han$^1$, Pradeep Kumar$^{1,2}$, and H. Eugene Stanley$^1$}
 \affiliation{$^1$Center for Polymer Studies and Department of Physics,
 Boston University, Boston, MA 02215\\ 
              $^2$Center for Studies in Physics and Biology, The
 Rockefeller University, New York, NY 10021}

\date{\today --- hks.tex}

\begin{abstract}

We perform molecular dynamics simulations to investigate the diffusive
motion of TIP5P water in the direction perpendicular to the two
hydrophobic confining walls.  To calculate the diffusion coefficient,
we use the concept of the characteristic residence time which is
calculated from the exponential decay of the residence time
probability distribution function.  We find that a diffusion anomaly of
water, increase of diffusion upon compression, is absent in the
direction perpendicular to the confining walls down to the lowest
temperature we simulate, 220K, whereas there is a diffusion anomaly, similar to that in bulk water, in
the direction parallel to the walls.  The absence
of a diffusion anomaly may arise mainly due to 
nanoconfinement, rather than due to the hydrophobic property of the confining
walls.

\end{abstract}

\maketitle

In addition to the thermodynamic anomalies of water
\cite{pablo,gene,angell1,angell2}, there are the salient dynamic
anomalies.  For example, in contrast to simple liquids, where the diffusion
coefficient $D$ decreases upon compression, for water at low temperature $D$ first
increases upon compression until reaching a maximum and then decreases
upon further increase of pressure \cite{gene2,priel1,angell3,priel2}.

Confined water shows many structural and dynamic properties different
from bulk water due to the geometry of confinement and the interaction
with the confining walls
\cite{pradeep1,zangi1,zangi2,zangi3,koga1,koga2,koga3,nicolas,morishige,gallo1,slovak1,slovak2,gallo2,gallo3,pradeep2,zangi4,bergman, brovch}.
Studies of water confined between two hydrophobic plates suggest that
the diffusion coefficient $D_{||}$ along the direction \textit{parallel} to the
plates displays a diffusion anomaly, which is the existence of a maximum
of $D_{||}$ as a function of a density at constant temperature, similar
to bulk water \cite{pradeep1}.  Moreover, the
temperature below which the anomalous diffusive behavior occurs is
shifted to lower temperature by about 40 K compared to bulk water
\cite{pradeep1}.

Dynamics in the direction \textit{perpendicular} to the confining plates is
modified by the small distance between plates ($\sim 1$~nm).
Studies have been carried out to investigate the diffusive motion in the
confining direction for many confined systems
\cite{liu,sega,lin,benesch}.  Liu \textit{et al.} showed that a
different treatment for the diffusion coefficient is needed in the
confining direction and used a ``dual simulation method" to calculate
the more precise diffusion coefficient at a liquid-vapor interface
\cite{liu}.  Sega \textit{et al.}  pointed out two different time
regimes (100~ps and 1~ns) of the mean square displacements (MSD) for the
complete description of diffusion in confined water and showed a
nonlinear MSD due to both spatial inhomogeneity and confinement
\cite{sega}.  The diffusion coefficient $D_{\perp}$ in the direction
perpendicular to the walls is difficult to find precisely in
nanoconfined water due to the finite and very small space available for
water molecules.  Before reaching the diffusive regime at which MSD is
linearly proportional to time, MSD already enters the plateau regime 
due to displacement bounded by a finite nanosize space.  Hence, $D_{\perp}$ 
cannot be extracted from the MSD. The Green-Kubo relation
for $D_{\perp}$ is also not valid for the confined system, as mentioned
in Ref.~\cite{liu}.

Here we propose an alternative approach to calculate $D_{\perp}$ and ask 
whether the diffusion anomaly of water, which has been shown
to exist in the parallel direction and in bulk dynamics, also exists in
the direction perpendicular to the hydrophobic confining plates.

We perform molecular dynamics (MD) simulations of $N=512$ TIP5P
\cite{mahoney} water molecules confined between two infinite parallel
smooth plates, separated by 1.1~nm, which are able to contain 2$\sim$3
layers of water molecules.  The plates are located at the positions z=
$\pm 0.55$ nm.  We model the water-wall interaction by a 9-3
Lennard-Jones (LJ) potential which is commonly used to represent the
effective interaction of water molecules with the confining plates \cite{steele,hansen}.  We choose the
parameters for the confining potential to have the hydrophobic property,
as in Ref.~\cite{pradeep1}.  Details of the simulations are also
described in Ref.~\cite{pradeep1}.  We perform MD simulations for seven
temperatures $T=220$, 230, 240, 250, 260, 280, and 300~K and for eight
densities $\rho=0.80$, 0.88, 0.95, 1.02, 1.10, 1.17, 1.25, and
1.32~g/cm$^3$.  These densities are calculated by considering the
accessible space between the walls, as explained in
Ref.~\cite{pradeep1}.  Periodic boundary conditions are used in the $x$
and $y$ directions.

\begin{figure}
\begin{center}
\includegraphics[width=0.7 \textwidth]{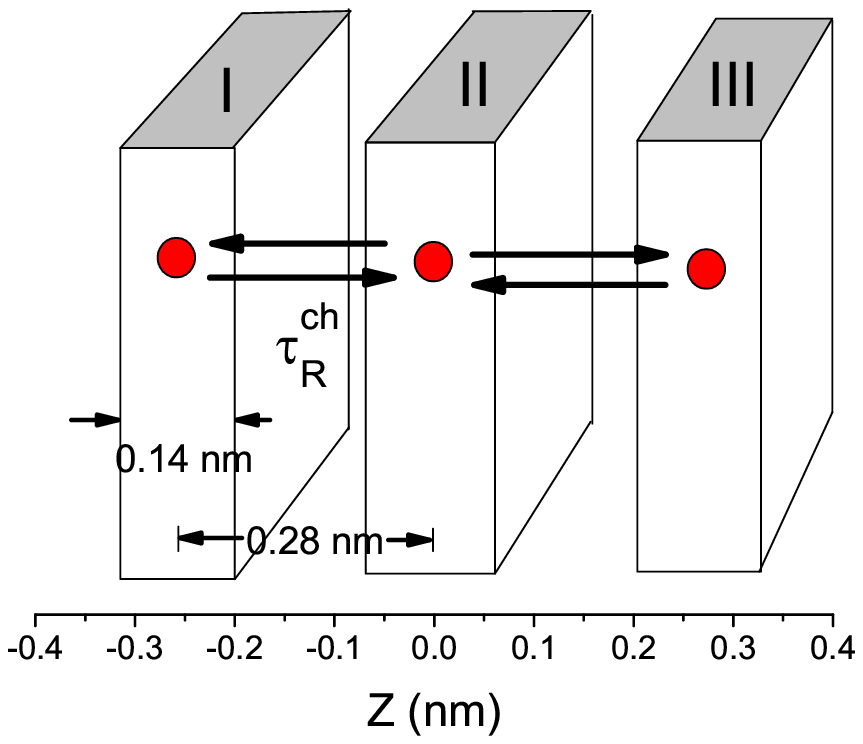}
\caption{\label{fig:region}(color online). Schematic description of the
  motion of water molecules in confined space with three defined
  residence regions. Two confining walls are located at $z=\pm 0.55$nm. The red circle represents a water molecule. We
  define the size of one residence region as 0.14 nm and the separation
  between two regions as 0.28 nm. $\tau^{ch}_R$ denotes the characteristic
  residence time calculated from the residence time distribution. On
  average, water molecules diffuse the distance of the separation between regions,
  $R_z=0.28$~nm, in the $z$ direction perpendicular to
  the confining walls during a time $\tau^{ch}_R$.}
\end{center}
\end{figure}

To calculate $D_{\perp}$, we first divide the system into three residence regions along $z$
direction such that there are two symmetric adjacent regions to the
surface and one middle region (see Fig.~\ref{fig:region}).  The width of
each region is 0.14~nm, and the separation between two adjacent regions
is $R_z=0.28$~nm which is the same as the linear size of a water
molecule.  Then, we calculate the residence time distributions
$P(\tau_R)$ of water molecules in the given residence region.  The
residence time is defined as the time over which water molecules stay in
one region before leaving it.

\begin{figure}
\begin{center}
\includegraphics[width=0.7 \textwidth]{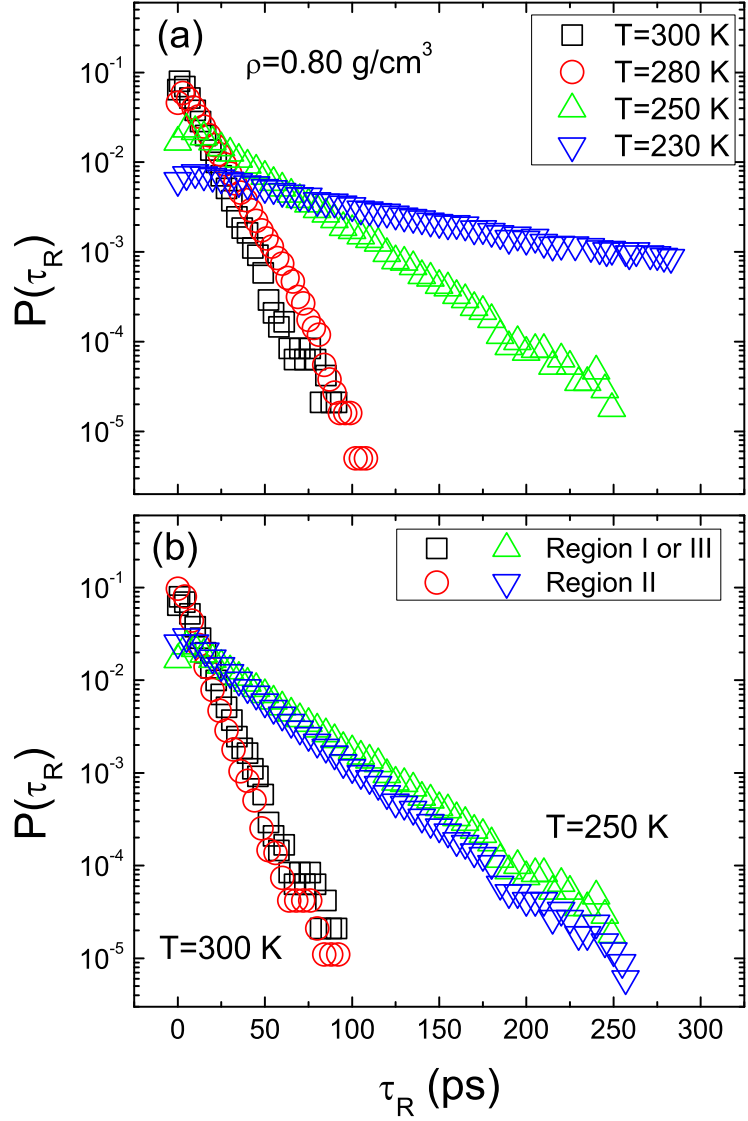}
\caption{\label{fig:rtd}(color online). (a) A semi-log plot of the
  residence time distribution function $P(\tau_R)$ of TIP5P water
  molecules at several temperatures and density
  $\rho=0.80$~g/cm$^3$. For all temperatures and densities investigated here, 
  $P(\tau_R)$ decays exponentially [$\sim\exp (-
  \tau_R/\tau_R^{ch})$]. (b) The residence time distributions $P(\tau_R)$
  for different residence regions (see Fig.~\ref{fig:region}.) at $\rho=0.80$~g/cm$^3$ and $T=250$~K
  and 300~K.}
\end{center}
\end{figure}

In Fig.~\ref{fig:rtd}, we show that $P(\tau_R)$ of the hydrophobic
confined water decays exponentially for all temperature investigated.
It was shown that the anomalous dynamics of water becomes subdiffusive
when it is confined in Vycor pores with the condition of low hydration
\cite{gallo4}.  This subdiffusive motion is related to a power law
behavior of $P(\tau_R)$.  However, our results show that in hydrophobic
confinement, $P(\tau_R)$ has an exponential behavior for all temperatures and densities
investigated.  We calculate the characteristic residence time
$\tau_{R}^{ch}$ by finding the inverse slope of a straight line fit to
$P(\tau_R)$ on the semi-log plot
\begin{equation}
P(\tau_R) \sim \exp\left(-{\tau_R\over\tau_R^{ch}}\right).
\end{equation}  
On average, during $\tau_{R}^{ch}$ water molecules diffuse the same
distance as the separation of two defined regions, as shown in
Fig.~\ref{fig:region}.

Since water molecules, on average, diffuse the distance equal to the separation
between two regions in the time interval $\tau_R^{ch}$, we
can write down the diffusion coefficient $D_\perp$
\begin{equation}
D_\perp=\frac{\langle R_z^2\rangle}{2\langle\tau_R^{ch}\rangle},
\end{equation}
where $D_\perp$ is the diffusion coefficient in the direction
perpendicular to the walls, $\langle R_z\rangle$ denotes the separation
between two residence regions, and $\langle\tau_R^{ch}\rangle$ is the
characteristic residence time averaged over an ensemble.  To get the
average value of $D_{\perp}$, we use the characteristic residence times
averaged over three different residence regions.  As shown in
Fig.~\ref{fig:rtd}(b), $P(\tau_{R})$ has the same exponentially decaying
behavior in different residence regions for all temperatures and densities
investigated.

\begin{figure}
\begin{center}
\includegraphics[width=0.7 \textwidth]{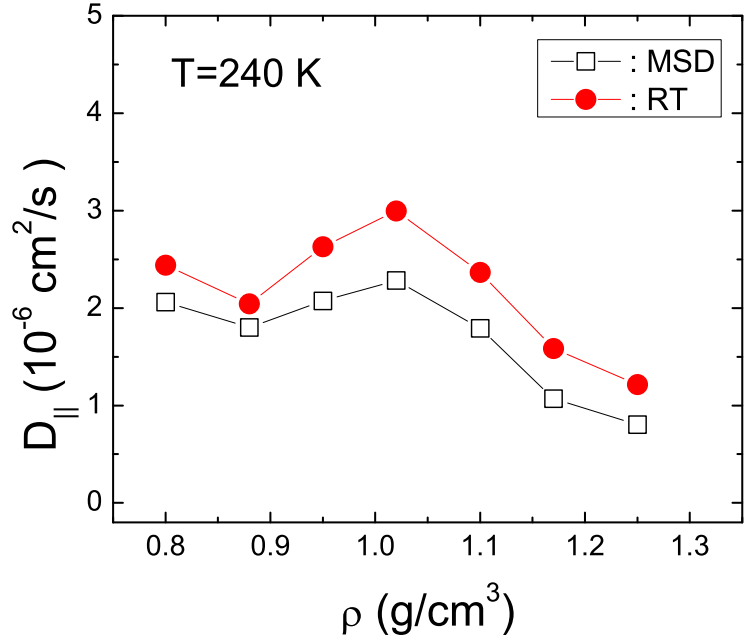}
\caption{\label{fig:diffpara}(color online). The diffusion coefficient $D_{||}$ 
in the direction parallel to the confining walls as a function of density. $D_{||}$ is calculated by 
using both from the Einstein relation (MSD) and the charateristic residence time (RT) 
at T=240 K. Both calculations exhibit a diffusion anomaly in $D_{||}$ as a function of density with a maximum
 at $\rho$ = 1.02 $g/cm^3$.}
\end{center}
\end{figure}

Since the calculation of $D_{||}$ is possible from the Einstein relation \cite{rapaport}, 
we investigate the validity of Eq. (2) by calculating $D_{||}$ 
using both the Einstein relation and the characteristic residence time.
In Fig.~\ref{fig:diffpara}, we show $D_{||}$ as a function of density for T=240 K calculated from both methods for a comparison.
We find that the method using the characteristic residence time gives 
slightly larger value of $D_{||}$ than using MSD, as shown in Fig.~\ref{fig:diffpara}.
Both methods for $D_{||}$ in our simulations exhibit a diffusion anomaly 
with a maximum at $\rho$ = 1.02 $g/cm^3$ as a function of density,
 same as in Ref. \cite{pradeep1}.
Therefore, we can say that Eq. (2) gives the correct value of $D$
 and can be used for the calculation of $D_{\perp}$ to investigate the existence of
a diffusion anomaly in the perpendicular direction.

\begin{figure}
\begin{center}
\includegraphics[width=0.7 \textwidth]{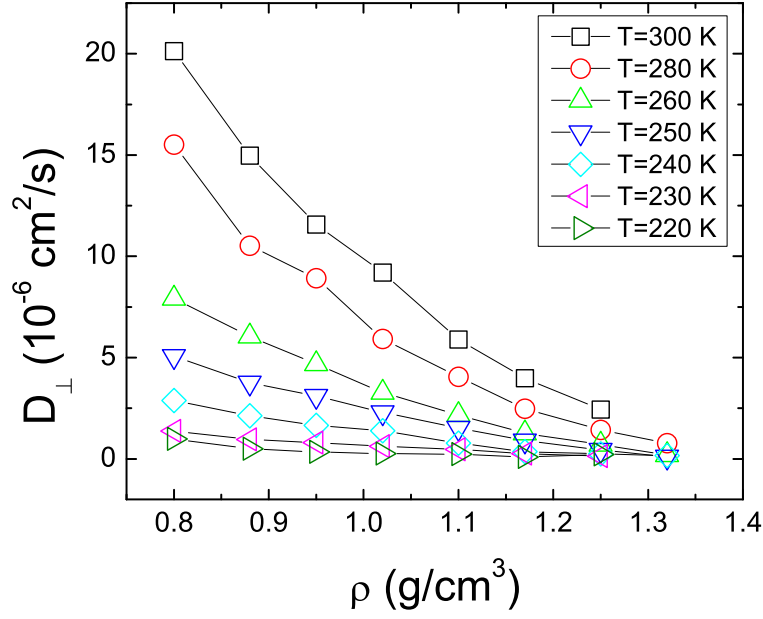}
\caption{\label{fig:diffconrho}(color online). The diffusion coefficient
  $D_{\perp}$ in the perpendicular direction as a function of density
  along constant temperature paths. $D_{\perp}$ decreases as density
  increases over the entire temperature range investigated. Our results
  suggest that there is no diffusion anomaly along the perpendicular
  direction even at the lowest temperature T=220 K we simulated. In the
  parallel direction, however, there is a diffusion anomaly at
  temperature lower than $T=250$~K. (see Fig.~\ref{fig:diffpara} and Ref.~\cite{pradeep1}.)}
\end{center}
\end{figure}

In Fig.~\ref{fig:diffconrho}, we show $D_{\perp}$ as a function of
density for all temperatures studied.  $D_{\perp}$ decreases as the
density increases.  Contrary to the diffusion anomaly found in the
parallel direction (similar to bulk water), as shown in Fig.~\ref{fig:diffpara}, 
our results show a diffusion
anomaly in the perpendicular direction does not exist down to the lowest
temperature we simulated.  As a result, we conclude that 
a diffusion anomaly of water is absent in the confining
direction down to very low temperatures in nanoconfinement.  From the fact that a diffusion anomaly
exists in the parallel direction but not in the perpendicular direction,
the main contribution to the absence of a diffusion anomaly in the
perpendicular direction might be the nanoconfinement rather than
the hydrophobic property of the confining walls.

\begin{figure}
\begin{center}
\includegraphics[width=0.7 \textwidth]{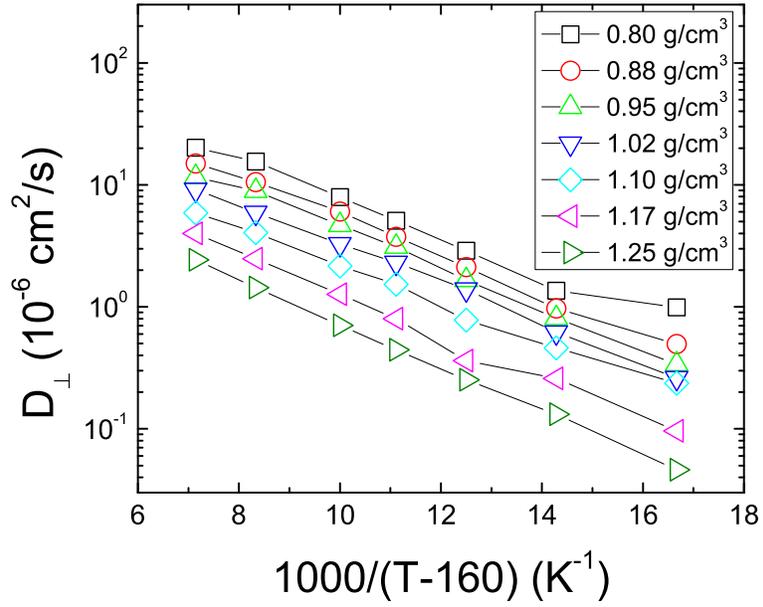}
\caption{\label{fig:diffcontem}(color online). A semi-log plot of
  $D_{\perp}$ as a function of $1/(T-T_0)$ along the constant density
  paths. The temperature dependence of $D_{\perp}$ can be fit with a
  Vogel-Fulcher-Tammann (VFT) form using the parameter value of
  $T_0=160$~K.}
\end{center}
\end{figure}

Next we study $D_{\perp}$ as a function of temperature along a constant
density path.  In Fig.~\ref{fig:diffcontem}, we find the temperature
dependence of $D_{\perp}$ can be fit with a Vogel-Fulcher-Tammann (VFT)
form for all the densities studied,
\begin{equation}
D_\perp=D_\perp^0\exp\left(-\frac{A}{T-T_0}\right).
\end{equation}
Here $D_\perp^0$, $A$, and $T_0$ are fitting parameters, and we use 160
K as the value of the parameter $T_0$.  It has been experimentally
observed that there is a fragile-to-strong transition near T= 220 K in
both supercooled water confined in micellar templated mesoporous silica
matrices MCM-41 \cite{liu1} and DNA and protein hydration water \cite{chen}, which
is shown to be connected to a liquid-liquid phase transition scenario
\cite{poole,mish,brov1,xu,pradeep3,paschek}.  In computer simulations of TIP5P
hydration water at atmospheric presseure, this non-Arrhenius to
Arrhenius crossover occurs at $T \approx 250$~K, the temperature at which the
isobaric specific heat has a maximum \cite{xu, pradeep3}.  In contrast to
simulation results of TIP5P bulk water, we find that in the confining
direction the TIP5P water confined between two hydrophobic plates does
not show a non-Arrhenius to Arrhenius dynamic crossover down to the
lowest temperature we simulated, $T=220$~K, suggesting that if there is
a crossover in the dynamics it would occur at much lower temperature
compared to bulk water.  This finding is consistent with the temperature
shift found for thermodynamic and dynamic properties of water confined
between hydrophobic surfaces \cite{pradeep1,chu}.

In summary, we have performed MD simulations of TIP5P water to calculate
and investigate the self-diffusion coefficient in the direction
perpendicular to the hydrophobic confining walls.  We calculated the
diffusion coefficient using a form similar to the Einstein relation of
self-diffusion, described with the separation between two residence
regions and the characteristic residence time found from the
exponentially decaying residence time distribution.  Contrary to the
diffusive dynamics of bulk water and the dynamics of confined water in the parallel direction, our simulation
results show that a diffusion anomaly does not exist in the direction
perpendicular to the confining walls.  As the density increases, the
diffusion coefficient keeps decreasing without reaching a maximum over
the whole temperature range investigated.  By comparing the
perpendicular direction to the parallel direction, it is reasonable to
conclude that this absence of a diffusion anomaly may arise due to
nanoconfinement rather than the hydrophobic property of the confining
walls.  In addition, we find that the temperature dependence of the diffusion
coefficient along the constant density path in the perpendicular
direction shows a VFT form.

%\begin{acknowledgments}

We thank C. A. Angell, J. A. Brunelle, S. V. Buldyrev, W.-S. Jung, J. Kim, J. Luo, M. G. Mazza,
K. Stokely, E. Strekalova, L. Xu, and Z. Yan for the helpful
discussions, and NSF grant CHE0606489 for support.

%\end{acknowledgments}

\newpage

\end{document}